\documentclass[aip,jcp,11pt,citeautoscript,reprint]{revtex4-1}

\pdfoutput=1

\newcommand{\figurewidth}{0.46\textwidth}

\usepackage{graphicx}
\usepackage{comment}

\begin{document}

\title{Monte Carlo cluster algorithm for fluid phase transitions in
  highly size-asymmetrical binary mixtures}

\author{Douglas J. Ashton}
\affiliation{Department of Physics, University of Bath, Bath BA2 7AY,
  United Kingdom}
\author{Jiwen Liu}
\affiliation{ Department of Materials Science and Engineering,
  University of Illinois at Urbana-Champaign, Urbana, Illinois 61801,
  U.S.A.}
\author{Erik Luijten}
\affiliation{Department of Materials Science and Engineering \&
  Department of Engineering Sciences and Applied Mathematics,
  Northwestern University, 2220 Campus Drive, Evanston, IL 60208,
  U.S.A.}
\author{Nigel B. Wilding}
\affiliation{Department of Physics, University of Bath, Bath BA2 7AY,
  United Kingdom}

\begin{abstract}
  Highly size-asymmetrical fluid mixtures arise in a variety of physical
  contexts, notably in suspensions of colloidal particles to which much
  smaller particles have been added in the form of polymers or
  nanoparticles. Conventional schemes for simulating models of such systems
  are hamstrung by the difficulty of relaxing the large species in the
  presence of the small one. Here we describe how the rejection-free
  geometrical cluster algorithm (GCA) of Liu and Luijten [Phys.\ Rev.\ Lett.\
  \textbf{92}, 035504 (2004)] can be embedded within a restricted Gibbs
  ensemble to facilitate efficient and accurate studies of fluid phase
  behavior of highly size-asymmetrical mixtures.  After providing a detailed
  description of the algorithm, we summarize the bespoke analysis techniques
  of Ashton \emph{et al.} [J. Chem.\ Phys.\ \textbf{132}, 074111 (2010)] that
  permit accurate estimates of coexisting densities and critical-point
  parameters. We apply our methods to study the liquid--vapor phase diagram of
  a particular mixture of Lennard-Jones particles having a $10$:$1$ size
  ratio. As the reservoir volume fraction of small particles is increased in
  the range $0$--$5\%$, the critical temperature decreases by approximately
  $50\%$, while the critical density drops by some $30\%$. These trends imply
  that in our system, adding small particles \emph{decreases} the net
  attraction between large particles, a situation that contrasts with
  hard-sphere mixtures where an attractive depletion force occurs.
\end{abstract}

\maketitle
  
\section{Introduction and background}
\label{sec:intro}

Colloidal suspensions are a class of complex fluids that comprises systems as
diverse as protein solutions, liquid crystals, and blood.  Technologically,
colloidal suspensions feature in applications such as coatings, precursors to
advanced materials, and drug carriers\cite{russel1989}. One of the key issues
in all these systems is the phase behavior of the suspension, or more
generally its stability.  Attractive dispersion forces exist between uncharged
colloids that can engender phase separation or irreversible aggregation
resulting in a gel---undesirable features in many applications. Accordingly,
one seeks to control the phase behavior (as well as dynamical properties such
as the rheology~\cite{larson1999}) by modifying the form of the effective
interactions between the colloidal particles. There are several routes to
achieving this, including charge stabilization (via modification of the
\emph{p}H) and steric stabilization (via grafting of flexible polymers onto
the colloidal surface)\cite{HUNTER2001,larson1999}.  Alternatively, the
effective interactions, and hence colloidal phase behavior, may be manipulated
through the addition of nanoparticles, with nanoparticle size, concentration,
and charge as control parameters\cite{BELLONI00,Tohver01,LiuetalJCP2005}. The
simplest and most celebrated example concerns colloids which interact (to a
good approximation) as hard spheres. Adding nanoparticles in the form of small
nonadsorbing polymers engenders an \emph{attractive} ``depletion'' force
between the colloidal particles\cite{Asakura1954}. This attraction can drive
phase separation resulting in a colloid-rich (`liquid') phase and a
colloidal-poor (`gas') phase\cite{POON02}---a phenomenon akin to the
fluid--fluid transitions occurring in molecular liquids and their mixtures.
Yet richer behavior occurs when one transcends simple hard-sphere potentials
between the nanoparticles and the colloids.  For example, if the nanoparticles
are weakly attracted to the colloids but repel one another, they can form a
diffuse (nonadsorbed) ``halo'' around each colloid
particle\cite{Tohver01,LIU2004_1,KARANIKAS04,Liu2005_1,CHAN05,BARR06}. The net
influence on the effective colloid--colloid interaction depends on the
nanoparticle density in a nontrivial way\cite{MARTINEZ05,Liu2005_1,BARR06}.

In view of the broad range of effects that can arise when nanoparticles are
added to a colloidal suspension, their prototypical model representation,
namely a size-asymmetric fluid mixture, has attracted considerable theoretical
and computational attention over the past years.  Analytical approaches
typically either focus on drastically simplified
models\cite{Asakura1954,russel1989} or attempt to render the size asymmetry
tractable by integrating out the degrees of freedom associated with the small
(nano)particles (see Ref.~\onlinecite{DIJKSTRA99} for a review).  The latter
strategy yields a one-component system of colloids described by an effective
pair potential representing the net influence of the small particles.  One
shortcoming of this approach is that, for all but the simplest types of
nanoparticles, the mapping to a one-component system is approximate, because
it neglects many-body colloidal interactions that can considerably alter the
nanoparticle distributions and hence the interactions induced by them.  These
effects may be significant in the regimes of density at which phase separation
occurs\cite{AMOKRANE03}.  Recent work has additionally raised concerns
regarding the accuracy of effective potentials in this
regime\cite{HERRING07,Oettel2009}, and has also emphasized the significance of
corrections to the entropic depletion picture in real
colloids\cite{GERMAIN04,BECHINGER99} as well as the importance of
polydispersity\cite{wilding2005b,largo2006}.  On the other hand, various
computational techniques, most notably Monte Carlo (MC) methods, are capable
of explicitly incorporating fluctuation and correlation effects.  Conventional
MC techniques (which attempt to displace or insert and delete particles) are
restricted to fluids mixtures in which the size ratio is of order unity,
rendering them unsuitable for the simulation of colloid--nanoparticle
suspensions, where typical size ratios encountered can extend to one or two
orders of magnitude\cite{Tohver01}.  The computational bottleneck results from
the effective ``jamming'' of the large species by even low volume fractions of
the small particles.  However, this problem has been resolved by means of the
geometric cluster algorithm (GCA) of Liu and Luijten\cite{LIU2004_0,Liu2005},
in which configuration space is sampled via rejection-free collective particle
updates, each of which facilitates the large-scale movement of a substantial
subgroup of particles (a ``cluster'').  Although the original algorithm
operates in the canonical ensemble, and hence cannot address phase separation
phenomena directly, in previous work\cite{LIU2006} we have developed a
generalization that embeds the GCA in the restricted Gibbs ensemble~(RGE),
such that clusters containing both large and small particles are exchanged
between two simulation boxes of fixed equal volumes. The resulting density
fluctuations within one box can be analyzed to determine the phase behavior.

The purpose of the present paper is firstly to provide a more detailed
description of the basic GCA--RGE algorithm that was introduced in
Ref.~\onlinecite{LIU2006}, and to integrate recent advances that we have made
in data analysis methods for determining coexistence and critical-point
properties within the RGE\cite{Ashton2010}. We then apply the improved
methodology to study the liquid--vapor coexistence properties of a mixture of
Lennard-Jones (LJ) particles having a size ratio $q=0.1$. In so doing we adopt
one aspect of effective fluid approaches, namely we focus on the liquid--gas
phase coexistence properties of the \emph{large} species (colloids) which are
assumed to be immersed in a supercritical fluid of small particles of
quasi-homogeneous density. This choice of perspective mirrors the experimental
reality, namely that often only the colloidal particles can be individually
imaged.  Accordingly, the phase diagrams that we present are single-component
projections (i.e., referring to the large species) of the full phase diagram,
obtained at a prescribed reservoir volume fraction of the small species, which
we vary in the range $0$--$5\%$. Note, however, that the small particles are
treated \emph{explicitly} and \emph{exactly} in our simulations, making this
method superior to effective-potential approaches which integrate out the
degrees of freedom associated with the small particles.

This paper is arranged as follows. Section~\ref{sec:model} introduces our
model system, a binary Lennard-Jones fluid. The GCA--RGE MC algorithm capable
of simulating this system in the highly size-asymmetrical limit is described
in Sec.~\ref{sec:method} together with an outline of techniques for
determining phase coexistence properties and critical-point parameters within
the RGE\@. Moving on to our results, Sec.~\ref{sec:results} presents
measurements of the large-particle coexistence densities as a function of the
reservoir volume fraction of small particles. We also discuss the underlying
reasons for the observed trends in the coexistence properties in terms of
measurements of the fluid structure. Finally, Sec.~\ref{sec:discuss} considers
the implications of our findings, the efficiency of our simulation approach
compared to more traditional schemes, and an outlook for further work.

\section{Model system}
\label{sec:model}

The model with which we shall be concerned is a binary mixture of
spherical particles, whose two species are denoted $l$ (large) and $s$
(small). Pairs of particles labeled $i$ and $j$ (having respective
species labels $\gamma_i$ and $\gamma_j$) interact via a Lennard-Jones  
potential,
\begin{equation}
  \phi_{ij}(r) = 4 \varepsilon_{\gamma_i\gamma_j} \left[
    \left(\frac{\sigma_{\gamma_i\gamma_j}}{r}\right)^{12} -
    \left(\frac{\sigma_{\gamma_i\gamma_j}}{r}\right)^6\right] \;,
  \label{eq:lj}
\end{equation}
where $\varepsilon_{\gamma_i\gamma_j}$ is the well depth of the interaction
and $\sigma_{\gamma_i\gamma_j}$ sets its range based on the additive mixing
rule $\sigma_{\gamma_i\gamma_j}=(\sigma_{\gamma_i}+\sigma_{\gamma_j})/2$.
$\sigma_{\gamma_i}$ and $\sigma_{\gamma_j}$ represent the particle
diameters. Interactions are truncated at $r_c=2.5\sigma_{\gamma_i\gamma_j}$
and we take $\sigma_l$ as our unit length scale.

In Sec.~\ref{sec:results} we study the case $q\equiv\sigma_{ss} /
\sigma_{ll}=0.1$, i.e., a $10$:$1$ size ratio.  We shall determine the phase
coexistence properties of the large particles as a function of temperature for
a prescribed reservoir volume fraction~$\eta_s^r$ of small particles, as
controlled by the imposed chemical potential of small particles~$\mu_s$.  It
should be noted, however, that since the small particles are not infinitely
repulsive, their volume fraction is notional in the sense that we use the
value of $\sigma_s$ as if it were a hard-core radius, i.e., we take
$\eta_s^r=\pi \bar{N}_s\sigma_s^3/(6V)$ where $\bar{N}_s$ is the average of
the fluctuating number of small particles contained within the system
volume~$V$.

Since we adopt the viewpoint that the small particles act as a background to
the large ones, we set $\varepsilon_{ss} = \varepsilon_{ls} =
\varepsilon_{ll}/10$, which ensures that the small-particle reservoir fluid is
supercritical in the temperature range of interest here, namely down to well
below the critical point of the large particles.  It is therefore natural to
define the dimensionless temperature~$T$ in terms of the well depth of the
interaction between the large particles, i.e.,
$T=k_{\rm B}\tilde{T}/\varepsilon_{ll}$, where $k_{\rm B}$ is Boltzmann's
constant and $\tilde{T}$ the absolute temperature.

\section{Methodology}
\label{sec:method}

\subsection{The GCA--RGE algorithm}
\label{sec:GCA--RGE}

In the original GCA\cite{LIU2004_0}, a fixed number of particles is located in
a single, periodically replicated simulation box of volume~$V$.  These
particles are then moved around via cluster moves, in which a subset of the
particles (identified by means of a probabilistic criterion) is displaced via
a geometric symmetry operation.  To realize density fluctuations, we employ
two simulation boxes and exchange particles between both boxes, as in the
Gibbs ensemble\cite{PANAGIO87}.  However, rather than exchanging individual
particles, we use the GCA to exchange entire \emph{clusters} of particles, so
that we retain the primary advantage of the GCA, namely the rapid
decorrelation of size-asymmetric mixtures.  As in the original GCA, a variety
of symmetry operations is possible; to connect to the original
description\cite{LIU2004_0}, we phrase the algorithm here in terms of point
reflections with respect to a pivot.  Since a point reflection will generally
displace some particles outside of the original simulation cell, we need to
adopt periodic boundary conditions for both simulation cells. Moreover, as
will transpire below, all particles that belong to a cluster and that are part
of the same simulation cell will retain their relative positions during the
cluster move.  Thus, the two simulation cells must have the same dimensions.
This symmetric choice, in which both cells have an identical, constant
volume~$V$, is referred to as the RGE\@.

We first describe the GCA--RGE for the case of a single species of particles
that interact through an isotropic pair potential~$V(r)$.  $N_0 = N_1 + N_2$
particles are distributed over the two simulation cells, with $N_1$ particles
in simulation cell~1 and $N_2$ particles in simulation cell~2. $N_0$ is chosen
to match a desired average density $\rho_0 = N_0/(2V)$.  A cluster move within
the GCA--RGE proceeds as follows.  A pivot is chosen at a random position
within simulation cell~1 and a second pivot is placed at the corresponding
position within simulation cell~2.  One of the $N_0$ particles is chosen as
the seed particle of the cluster.  This particle~$i$, which thus can be
located in either simulation cell, is point-reflected with respect to the
pivot (in its own simulation cell) from its original position~$\mathbf{r}_i$
to the new position~$\mathbf{r}_i'$.  However, rather than placing the
particle at the new position (modulo the periodic boundary conditions) in its
original box, we place it at the corresponding position $\mathbf{\bar{r}}_i'$
in the \emph{other} box.  Subsequently, in keeping with the methodology of the
GCA, all particles in the first box that interact with particle~$i$ in its
original position (the ``departure site'')~$\mathbf{r}_i$ as well as all
particles in the second box that interact with particle~$i$ in its new
position (the ``destination site'')~$\mathbf{\bar{r}}_i'$ are considered for
point reflection with respect to the pivot point in their respective box and
subsequent transfer to the opposite box.  These particles, which we refer to
with the index~$j$, are point-reflected and transferred with probability
\begin{equation}
  p_{ij} =  \max[1 - \exp(-\beta \Delta_{ij}), 0] \;,
  \label{eq:bprob}
\end{equation}
where $\beta = 1/(k_{\rm B} T)$ and $\Delta_{ij} = -
V(|\mathbf{r}_i-\mathbf{r}_j|)$ if $i$ and~$j$ reside (prior to the transfer
of particle~$i$) in the same cell. If $i$ and~$j$ initially reside in
different cells (and hence do not interact prior to the transfer of
particle~$i$), $\Delta_{ij} = V(|\mathbf{\bar{r}}_i'-\mathbf{r}_j|)$. This
process is repeated iteratively, i.e., for each particle~$j$ that is
transferred to the opposite box, all neighbors that interact with~$j$ either
near its departure site or near its destination site, and that have not yet
been transferred in the present cluster step, are considered for point
reflection and transfer as well.  This process proceeds until there are no
more particles to be considered; all particles that are indeed point-reflected
and transferred are collectively referred to as the cluster. Observe that the
pair energy of all particles that are part of the cluster remains unchanged:
If two particles reside in the same simulation cell prior to the cluster
construction and both become part of the cluster, their separation remains
constant.  Likewise, if two particles reside in different cells prior to the
cluster construction and both are transferred, then their interaction energy
is zero before and after the cluster move.  The same holds true for the pair
interactions between all particles that are \emph{not} part of the cluster.
Thus, the total energy change induced by the cluster move originates from the
change in pairwise interactions between members of the cluster and particles
that are not part of the cluster.  In the terminology of
Ref.~\onlinecite{LIU2004_0} such ``bonds'' are either broken if a particle is
included in the cluster whereas a neighbor near its departure site is not, or
formed if a particle interacts with a neighbor near its destination site, and
this neighbor does not become part of the cluster.

Although the cluster formation process is probabilistic, we note that $p_{ij}$
only depends on the pair potential between particles $i$ and~$j$, rather than
on the \emph{total} energy change resulting from the displacement and transfer
of particle~$j$.  As a result, the cluster algorithm is self-tuning: Overlaps
of repulsive particles will be avoided and strongly bound particles tend to
stay together.  Indeed, owing to the choice of the \emph{bond probability}
$p_{ij}$, Eq.~(\ref{eq:bprob}), no further acceptance criterion needs to be
applied upon completion of the cluster, leading to a rejection-free algorithm
in which large numbers of particles are moved nonlocally.  The proof of
detailed balance is identical to that provided in Ref.~\onlinecite{Liu2005}
for the original GCA, where it was demonstrated that the ratio of the
probability of constructing a cluster in a given configuration~$X$ leading to
a configuration~$Y$ [the transition probability $T(X \to Y)$] and the reverse
transition probability $T(Y \to X)$ is the inverse of the ratio of Boltzmann
factors of the respective configurations.  The presence of \emph{two}
simulation cells simplifies rather than complicates the proof, just like
$\Delta_{ij}$ in Eq.~(\ref{eq:bprob}) is a special case of the original
expression\cite{LIU2004_0}, owing to the fact that two particles do not
interact if they reside in opposing boxes.

The generalization to multiple species is straightforward, and does not
lead to any conceptual changes in the algorithm. Indeed, the GCA shows
its primary advantages in the simulation of size-asymmetric mixtures, as
it realizes nonlocal moves without the usual decrease in acceptance
ratio\cite{LIU2004_0}. However, whereas there is no limitation on the
number of species, the overall volume fraction must be kept below a
threshold value.  Above this threshold, which is related to the
percolation threshold and depends on system composition and interaction
strengths between the particles\cite{Liu2005}, the cluster frequently
contains the majority of all particles.  This is detrimental to the
performance of the algorithm, as it is computationally expensive to
construct such clusters, whereas the configurational change in the
system is very small.  The existence of this threshold also necessitates
the use of an implicit solvent, as is common in the simulation of
colloidal suspensions.  Moreover, for reasons explained in detail in
Sec.~\ref{sec:smalls}, in our simulations of binary mixtures we combine
the geometric cluster moves with grand-canonical moves for the small
species.  It is important to emphasize that the different types of MC
moves are independent.  Thus, the small species fully participate in the
cluster construction process and the advantage of nonlocal
rejection-free moves is retained, yet the density of small particles in
both simulation cells is controlled by a chemical potential~$\mu_s$.

In Ref.~\onlinecite{Liu2005}, a number of technical improvements to the GCA
are described. These can all be applied to the GCA--RGE algorithm.  Most
notably, it is possible to decrease the average cluster size, and hence
increase the packing fractions that can be simulated efficiently, through
biased placement of the pivot.  Furthermore, for mixtures of particles with
large size disparities, the cluster construction process can be facilitated by
employing multiple subcell structures and corresponding neighbor
lists\cite{allentildesley87,Liu2005}.

Lastly, we note that, during the preparation of our original
work\cite{LIU2006}, Buhot\cite{Buhot2005} proposed an approach that has
significant similarities to the GCA--RGE method. His method also employs two
boxes of identical size, and exchanges clusters of particles.  However, rather
than the GCA\cite{LIU2004_0} he uses the original geometric algorithm of Dress
and Krauth\cite{Dress1995}, which is only applicable to hard spheres.
Moreover, for each point reflection it is decided at random whether a particle
is transferred to the opposing box or not.  If this decision were made only
once per cluster (i.e., upon selecting the seed particle of the cluster), this
would amount to an alternation of the GCA--RGE with regular GCA moves.  On the
other hand, if it is decided independently for each particle that is added to
the cluster, the average cluster size will be larger than in the GCA--RGE,
generally an undesirable situation.  The most important difference, however,
between, our approach and Ref.~\onlinecite{Buhot2005} is that the latter can
only be used for the idealized case of \emph{symmetric} binary mixtures, where
the critical composition is known \emph{a priori}.  By contrast, in our method
we employ the relationship between the RGE and the grand-canonical ensemble to
derive a prescription for locating the critical point and coexistence curve
for general binary mixtures.

\subsection{Locating phase coexistence and criticality in the RGE}
\label{sec:methods}

The absence of volume exchanges between both simulation boxes in the
symmetrical restricted Gibbs ensemble implies that, unlike for the full Gibbs
ensemble\cite{PANAGIO87}, there is no automatic pressure equality and hence no
guarantee that the measured particles densities are representative of
coexistence. In this section we outline how one can nevertheless extract
coexistence properties from RGE simulations without resorting to direct
measurements of pressure. A fuller account of the theoretical basis of the
methods we describe can be found in Ref.~\onlinecite{Ashton2010}.

\begin{figure}
\includegraphics[width=\figurewidth]{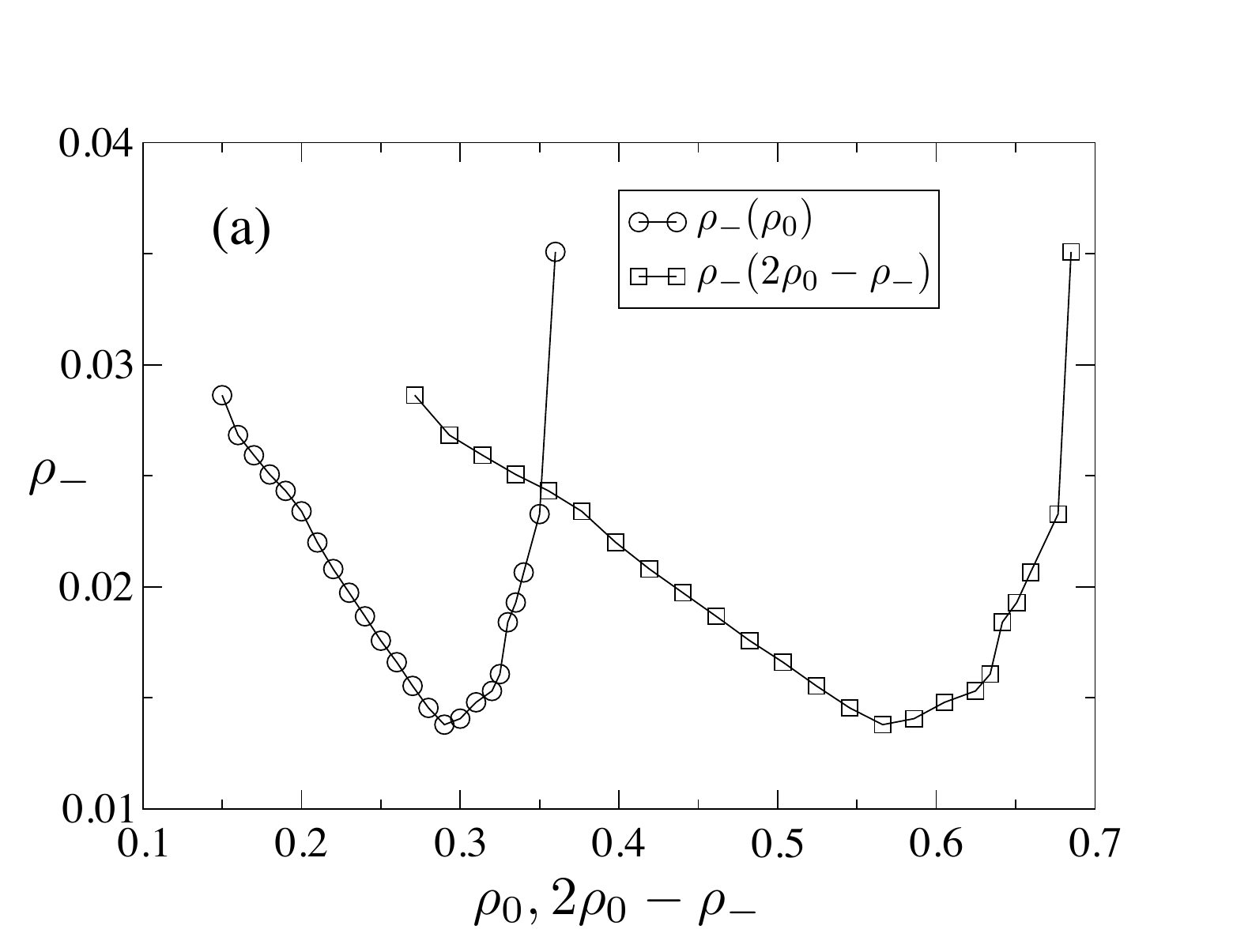} 
\includegraphics[width=\figurewidth]{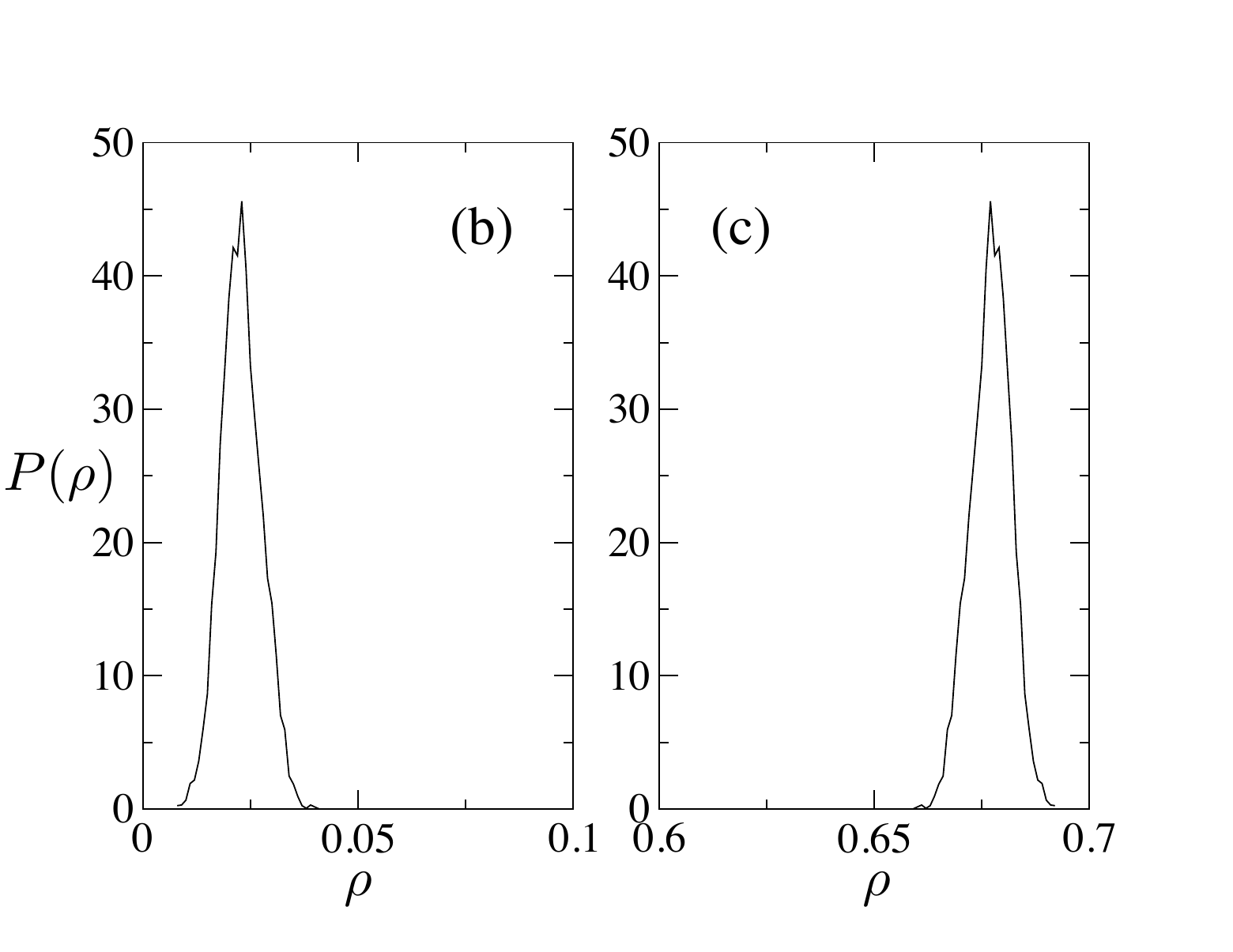} 
\caption{(a)~Illustration of the operation of the intersection method
  described in the text for the determination of the coexistence diameter
  density. Data is shown for the state point $\eta_s^r=0.01$, $T = 0.8
  (=0.764T_c)$. Plotted are measured estimates of the average value $\rho_-$
  of the low-density peak of $\hat{P}_L(\rho|\rho_0)$, for a series of
  values of~$\rho_0$.  The same data is also shown plotted against
  $2\rho_0-\rho_-$. The value of $\rho_0$ at which the two data sets intersect
  [$\rho_0=0.351(3)$] serves as an estimate of the coexistence diameter
  density~$\rho_d$. (b) and~(c): The measured peaks of $\hat{P}_L(\rho)$ for
  $\rho_0=\rho_d$, whose individual integrated averages yield estimates of the
  coexistence densities.}
\label{fig:intersection}  
\end{figure}

Within the RGE framework for our mixture, the total density~$\rho_0$ of
large particles across the two boxes is fixed. However, the one-box
density of large particles, $\rho \equiv N_1/V$, fluctuates. For any
given choice of $\rho_0$, the form of the probability distribution of
$\rho$, $\hat{P}_L(\rho)$, depends both on the temperature $T$ and on
the choice of the chemical potential~$\mu_s$ of the small particles. As
shown in Ref.~\onlinecite{Ashton2010}, measurement of the form of
$\hat{P}_L(\rho)$ for a range of values of $\rho_0$ provides a route to
the coexistence and critical-point parameters.  The basic strategy is as
follows. Within the RGE, one explores the coexistence region by varying
$\rho_0$ at fixed $T$ and~$\mu_s$. For $\rho_0$ sufficiently far inside
the coexistence region, the distribution $\hat{P}_L(\rho)$ exhibits a
double-peaked form, with peaks located at densities $\rho_-$ and~
$\rho_+$. In general, however, these peak densities do \emph{not}
coincide with the gas and liquid coexistence densities $\rho_{gas}$ and
$\rho_{liq}$---a situation which contrasts with the full Gibbs ensemble. An
important exception is when $\rho_0$ equals the coexistence diameter
density $\rho_d \equiv (\rho_{gas}+\rho_{liq})/2$, for which one
finds\cite{Ashton2010}: \begin{equation} \left. \begin{array}{ll} \rho_-
&= \rho_{gas}\\ \rho_+ &= \rho_{liq} \end{array} \right \} \hspace*{3mm}
\text{when $\rho_0=\rho_d$} \;. \label{eq:atdiameter} \end{equation}
Another important case is when $\rho_0=(\rho_{gas}+\rho_d)/2$ for which one
finds \begin{equation} \left. \begin{array}{ll} 
  \rho_- &= \rho_{gas}\\
  \rho_+ &= \rho_d
\end{array}
\right \} \hspace*{3mm} \text{when $\rho_0=(\rho_{gas}+\rho_d)/2$} \;.
\label{eq:atlowerdens}   
\end{equation}
Enforcing consistency between Eqs.\ (\ref{eq:atdiameter})
and~(\ref{eq:atlowerdens}) suffices to permit determination of $\rho_d$ and
hence [via Eq.~(\ref{eq:atdiameter})] the coexistence densities. It is
convenient to achieve this graphically (see Fig.~\ref{fig:intersection}) by
plotting the low density peak $\rho_-$ both against $\rho_0$ and against
$2\rho_0 - \rho_-$: The value of $\rho_0$ at which the two curves intersect
provides an estimate for the coexistence diameter $\rho_d$ and one can simply
read off the coexistence densities from the corresponding values of $\rho_-$
and~$\rho_+$. In Ref.~\onlinecite{Ashton2010} this ``intersection method'' was
shown to be very accurate for determining coexistence properties and to
exhibit finite-size effects comparable to those found in grand-canonical
simulations. Indeed it turns out to be much more accurate than the technique
we proposed previously for determining the coexistence diameter in the
RGE\cite{LIU2006}, wherein one determines $\rho_d$ as the value of $\rho_0$ at
which the variance of $\hat{P}_L(\rho)$ is maximized. We have found this
latter procedure to be considerably more sensitive to finite-size effects than
the intersection method, and it was therefore not used here.

Turning now to the matter of estimating critical parameters within the
RGE, an accurate technique for achieving this has been described in
detail in Ref.~\onlinecite{Ashton2010}. The basic idea is to measure the
``iso-$Q^\star$ curve'' introduced in Ref.~\onlinecite{LIU2006}, which
is simply the locus of points in $\rho_0$--$T$ space for which the
fourth-order cumulant ratio of $\hat{P}_L(\rho)$,

\begin{equation}
  Q \equiv \frac{\langle (\rho-\rho_0)^2\rangle^2}{\langle
    (\rho-\rho_0)^4\rangle}\:,
  \label{eq:qm0}
\end{equation}
matches the independently known\cite{LIU2006} fixed-point value
$Q^\star=0.711901$ appropriate to the Ising universality class and the RGE
ensemble.\footnote{Note that this definition of $Q$ only involves central
  moments because {$\hat{P}_L(\rho)$} is symmetric with respect to its mean
  {$\langle \rho\rangle=\rho_0$}.}

\begin{figure} 
  \includegraphics[width=\figurewidth]{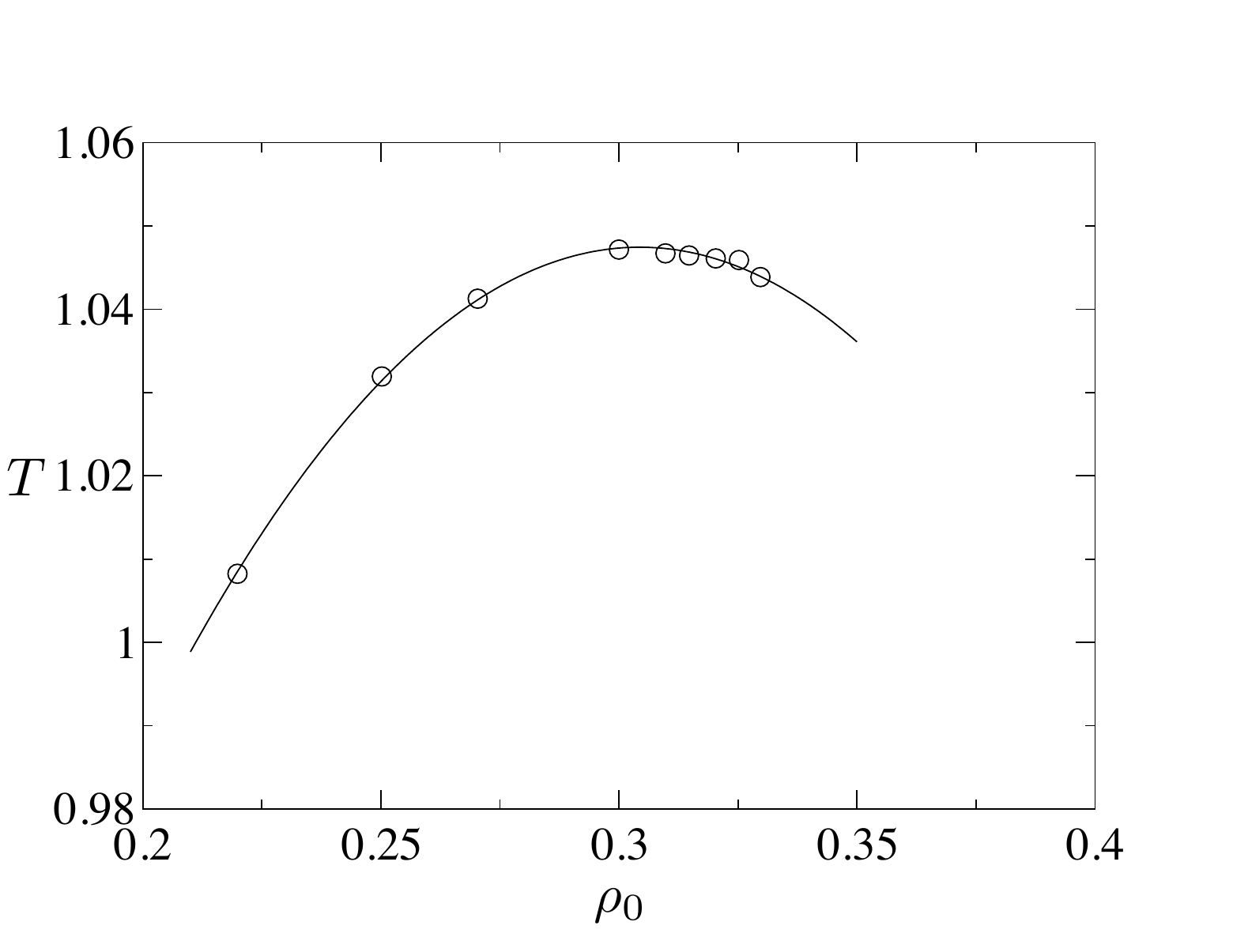} 
  \caption{Estimates of points on the iso-$Q^\star$ curve for $\eta_s^r=0.01$,
    obtained for a system of size $L=10$, as described in the text. A parabolic fit to the data (solid
    line) identifies the coordinates of the maximum of the curve which serves
    as a finite-size estimate of the critical-point parameters.}
  \label{fig:isoqstar}
\end{figure}

Now, it transpires\cite{LIU2006,Ashton2010} that the iso-$Q^\star$ curve is
essentially a parabola in $\rho_0$--$T$ space, the position of whose maximum
represents a finite-size estimator of the critical-point density~$\rho_c$ and
temperature~$T_c$. This maximum can be accurately located via a simple
quadratic fit to measured points on the iso-$Q^\star$ curve. In practice we
determine this curve as follows. A simulation is performed at some $\rho_0$
and $T$ to determine $\hat{P}(\rho)$. This distribution is then extrapolated
in temperature via histogram reweighting\cite{ferrenberg1989} to determine
that temperature for which $Q=Q^\star$. The procedure is then repeated for a
range of values of $\rho_0$ allowing us to trace out the whole iso-$Q^\star$
curve. An example of the resulting form of this curve is shown in
Fig.~\ref{fig:isoqstar}. Note that, in general, for reasons of computational
economy, the majority of the points that we determine on an iso-$Q^\star$
curve are for densities lower than the critical density, since the efficiency
of the cluster algorithm is greater at lower overall volume fractions of
particles.

Estimates of the critical parameters obtained from the iso-$Q^\star$
maxima for a range of system sizes can, in principle, be extrapolated to
the thermodynamic limit using finite-size scaling relations derived in
Ref.~\onlinecite{Ashton2010}, which fully account for both field-mixing
effects and corrections to scaling. Unfortunately, in the present work,
the computational cost of simulating more than one system size was found
to be prohibitive. However, the variations that we find in
critical-point parameters as a function of $\eta_s^r$ dwarf those that
one might expect on the basis of finite-size effects alone. Thus we are
nevertheless able to report reliable trends from our measurements.

\subsection{Treatment of the small particles}
\label{sec:smalls}

As was argued in Sec.~\ref{sec:intro}, it is the relaxation of the large
particles that constitutes the sampling bottleneck for highly
size-asymmetrical mixtures. Local MC updates of small particles are
computationally relatively unproblematic, irrespective of whether one performs
particle displacements or insertions and deletions.  Consequently, one has the
choice of treating the small particles canonically so that their density is
globally conserved, or grand canonically, in which case the density fluctuates
under the control of a prescribed chemical potential.  The choice one makes in
this regard greatly affects the manner in which the bulk phase behavior is
probed. Moreover it transpires that only the grand-canonical treatment of the
small particles is compatible with our intersection method
(Sec.~\ref{sec:methods}) for determining coexistence parameters.

\begin{figure}
  \includegraphics[width=\figurewidth]{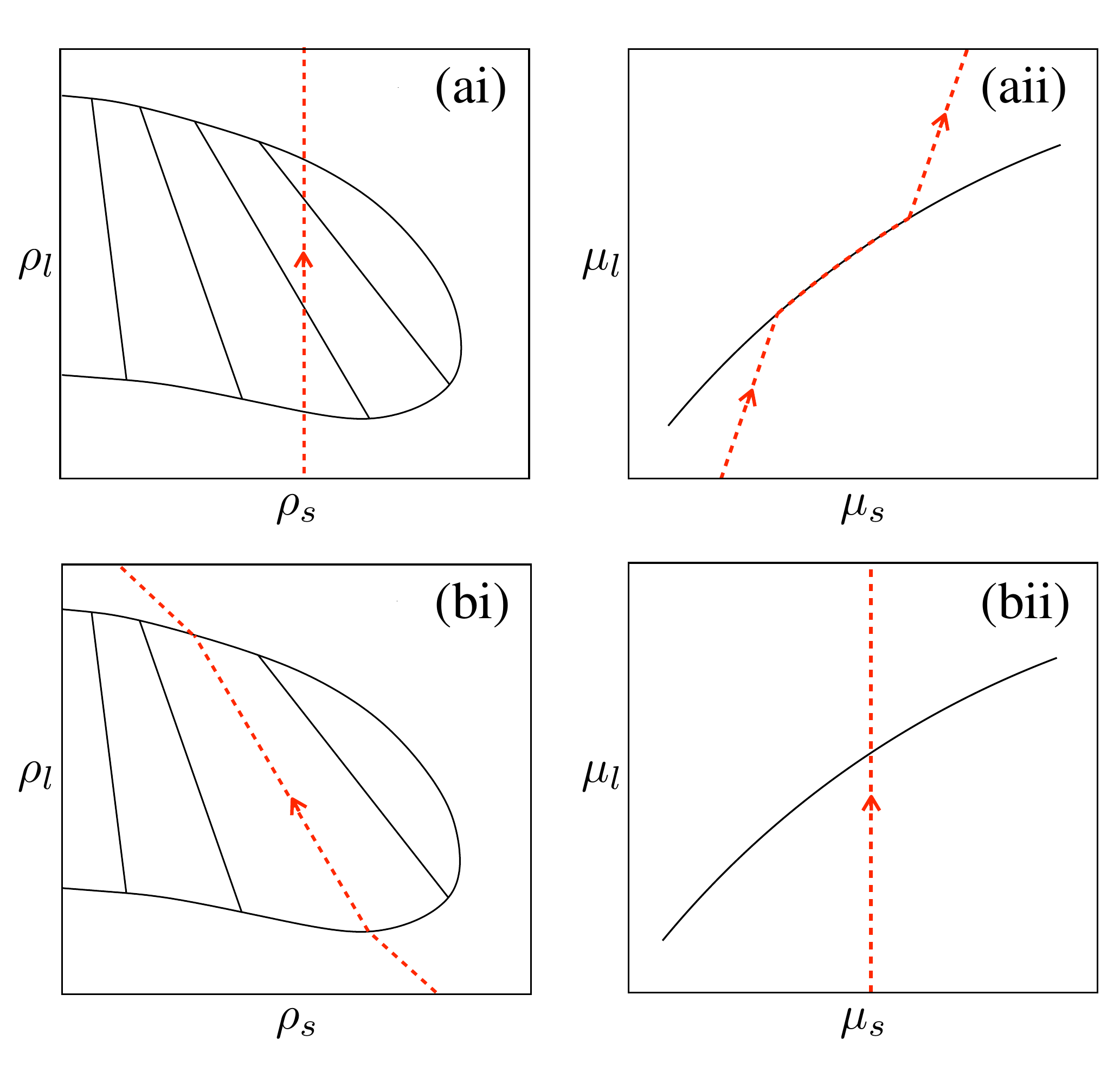} 
  \caption{Isothermal cuts through the exemplary phase diagram of a binary
    fluid mixture described in the text. Liquid--vapor coexistence is
    represented in terms of (i)~densities ($\rho_l$--$\rho_s$) and
    (ii)~chemical potentials ($\mu_l$--$\mu_s$). In (a) the coexistence region
    is crossed along a path of constant $\rho_s$, whereas in (b) it is crossed
    along a path of constant~$\mu_s$.}
  \label{fig:cut}
\end{figure}

To clarify these points, we show in Fig.~\ref{fig:cut} sketches of the
isothermal bulk phase diagram of an exemplary size-asymmetrical binary mixture
with large-particle density~$\rho_l$, small-particle density~$\rho_s$, and
conjugate chemical potentials $\mu_l$ and~$\mu_s$,
respectively. Figure~\ref{fig:cut}(ai) shows the phase behavior in the
$\rho_l$--$\rho_s$ plane, while the corresponding phase diagram in the
$\mu_l$--$\mu_s$ plane is shown in Fig.~\ref{fig:cut}(aii). In constructing
these sketches we have anticipated the behavior of the model of
Sec.~\ref{sec:model}, namely that the larger species has stronger attractive
interactions and thus phase separates on its own [vertical axis of
Fig.~\ref{fig:cut}(ai)] at the chosen temperature, while the small-particle
fluid (horizontal axis) does not. With interaction strengths chosen in this
way, the larger particles will typically accumulate in the liquid phase, with
its shorter interparticle distances, as shown by the representative tie lines
in the density representation.\footnote{In the case shown, the reduction in
  available volume for the small particles in the liquid compared to the gas
  leads to a reduction in their density relative to the gas. However, the sign
  of the gradient of the tie lines is expected to depend on the details of the
  large--small and small--small interaction strengths.} Note that in
chemical-potential space, coexistence occurs on a line of points, as shown in
Fig.~\ref{fig:cut}(aii).

Let us consider first the canonical scenario in which one traverses the
coexistence region at constant bulk density of small particles, as expressed
by the dashed trajectory included in Figs.\ \ref{fig:cut}(ai)
and~(aii). Clearly the tie lines in Fig.~\ref{fig:cut}(ai) cross any line of
constant $\rho_s$ moving from smaller values of $\rho_l/\rho_s$ at the gas end
to larger ones for the coexisting liquid. Hence this path generates a sequence
of pairs of coexistence states, one for each tie line crossed. The same
trajectory in terms of the chemical potentials $\mu_l$--$\mu_s$ is shown in
Fig.~\ref{fig:cut}(aii). Here the path followed first meets the coexistence
line, tracks along it it for some distance and then separates from it.

The alternative (grand-canonical) scenario, in which the small-particle
density is permitted to fluctuate at constant chemical potential~$\mu_s$, is
illustrated in Figs.\ \ref{fig:cut}(bi) and~(bii).  Here, as $\mu_l$ is varied
at constant $\mu_s$, the system crosses the coexistence line at a single
point. In density space the corresponding trajectory thus follows a particular
tie line though the coexistence region, as shown in Fig.~\ref{fig:cut}(bi).

In seeking to apply the RGE ensemble to study a binary mixture, it is
therefore imperative that one adopts a grand-canonical treatment of the small
particles. Doing so ensures that only a single pair of coexistence states is
encountered inside the bulk coexistence region, i.e., that one tracks a tie
line of the bulk phase diagram. This is a prerequisite for the correct
operation of our intersection method, which is designed to determine first the
diameter density for a single pair of coexistence states as a prelude to
determining the coexistence densities of the large particles themselves. We
further note that a grand-canonical treatment of the small particles
corresponds more closely to common experimental arrangements where one
typically measures properties of the mixture with respect to variations of a
reservoir volume fraction of small particles.

\section{Results}
\label{sec:results}

Equipped with the methods described above, we can set about the task of
determining the coexistence properties of the large particles in the presence
of a sea of small ones. To this end we apply the GCA--RGE method to study
liquid--vapor phase coexistence in a $q=0.1$ LJ mixture (see
Sec.~\ref{sec:model}). In addition to the cluster updates which swap whole
groups of particles (including both large and small species) between boxes,
small particles are sampled across both boxes using a standard local
grand-canonical algorithm at constant chemical potential (see
Sec.~\ref{sec:smalls}). As discussed above, we choose $\mu_s$ to yield (for
each temperature of interest) a prescribed volume fraction $\eta^r_s$ of small
particles in the reservoir.  This requires prior knowledge of the reservoir
equation of state $\eta_s^r(\mu_s,T)$, which we obtained via explicit
simulation of the pure fluid of small particles. Note that the computational
cost of obtaining $\eta_s^r(\mu_s,T)$ is low, particularly if one employs
histogram extrapolation\cite{ferrenberg1989} to scan a region of $\mu$ and $T$
surrounding each simulation state point.

\begin{figure}
  \includegraphics[width=\figurewidth]{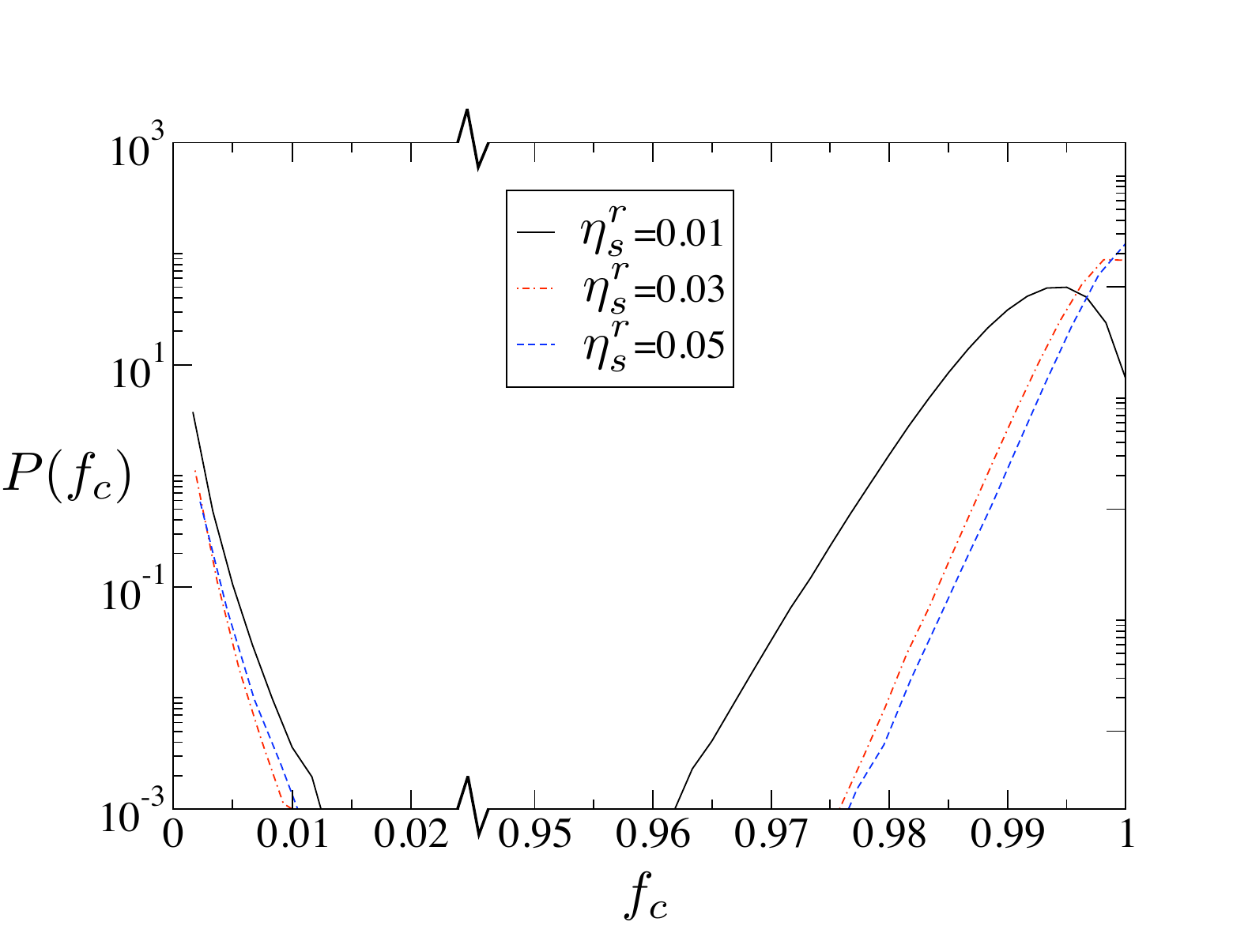} 
  \caption{Distribution of the fraction of large particles in the cluster, $f_c$, at
    criticality for the values of $\eta_s^r$ shown in the legend. }
\label{fig:clustersize}  
\end{figure}

The GCA--RGE simulations are performed using two cubic periodic
simulation boxes of linear size~$L=10$. We consider seven values of the
reservoir volume fraction of the small particles, $\eta^r_s=0.005$,
$0.01$, $0.015$, $0.02$, $0.03$, $0.04$, and~$0.05$.  In the limit of
low densities of large particles, these values of $\eta^r_s$ correspond
to average numbers of small particles in the range $10^4$--$10^5$. The
computational expenditure incurred in simulating such great numbers of
small particles places an upper bound on the value of $\eta_s^r$ for
which it is feasible to perform a full determination of the coexistence
binodal. Further difficulties arise from the fact that the typical
cluster size at coexistence was found to grow steadily as we increased
$\eta_s^r$. This is demonstrated in Fig.~\ref{fig:clustersize} which
plots the distribution of the fraction of large particles in the cluster
for $\eta_s^r=0.01, 0.03$ and $0.05$ measured at the respective critical
point parameters.  These distributions are bimodal, with some fairly
small clusters comprising just a few particles, and many clusters that
comprise the vast proportion of large particles. Updating such clusters
results in only relatively minor alterations to a configuration and
consequently, we are able to determine the coexistence binodal only for
$\eta_s^r\le 3\%$, whereas for $\eta_s^r=0.04$ and $0.05$ we restrict
ourselves to determining critical-point parameters. It should be
stressed however, that the cluster sizes observed in the present study
may not provide a general guide to the maximum $\eta_s^r$ at which the
GCA-RGE scheme will operate. This is because, as we shall show, our
choice of interspecies interactions engenders a large depression in
$T_c$ with increasing $\eta_s^r$ which in turn promotes the formation of
large clusters due to the temperature dependence of the GCA bond-formation
probability Eq.~(\ref{eq:bprob}). However, other choices of interactions
can be expected to lead to a different temperature dependence of the critical point
parameters, hence allowing larger values of $\eta_s^r$ to be attained.

\begin{figure}
  \includegraphics[width=\figurewidth]{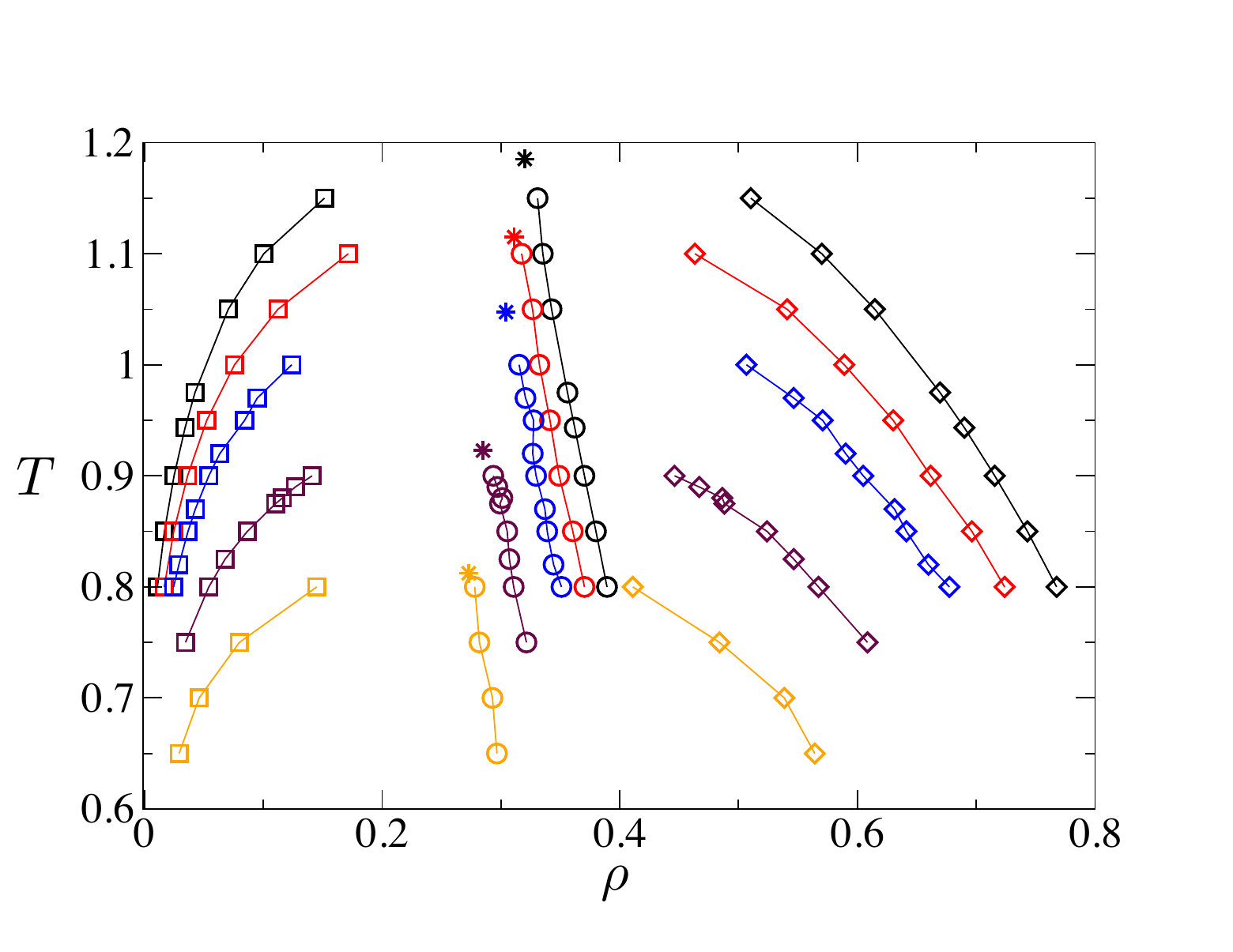} 
  \caption{Phase diagrams showing the liquid (diamonds) and gas (squares)
    coexistence densities of large particles for $q=0.1$. Data are shown for
    reservoir volume fractions (top to bottom) $\eta_s^r=0$, $0.005$, $0.01$,
    $0.02$, and~$0.03$. Also shown in each case is the coexistence diameter
    (circles) and critical point (asterisks).}
  \label{fig:phasediagram}
\end{figure}

\begin{figure}
  \includegraphics[width=\figurewidth]{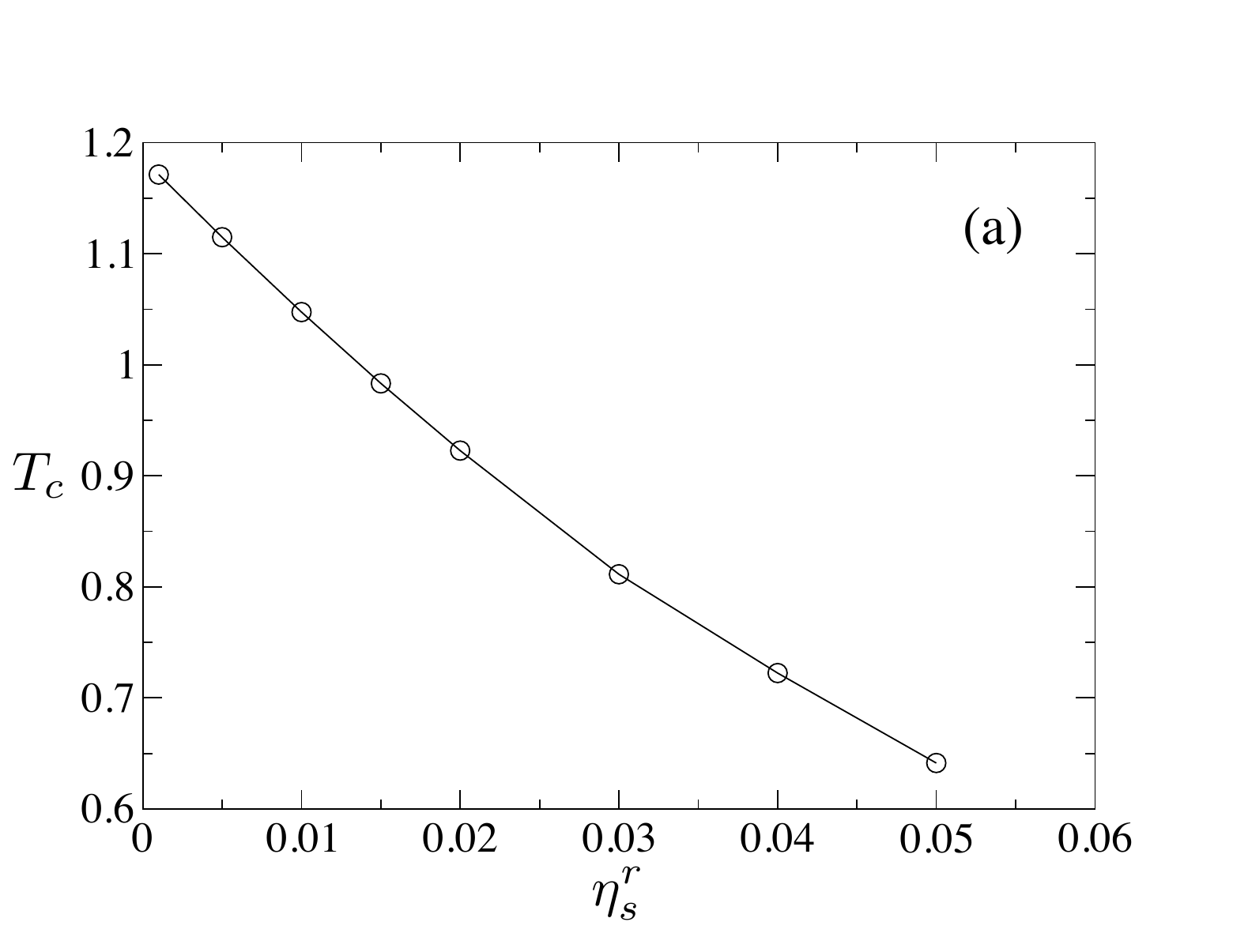} 
  \includegraphics[width=\figurewidth]{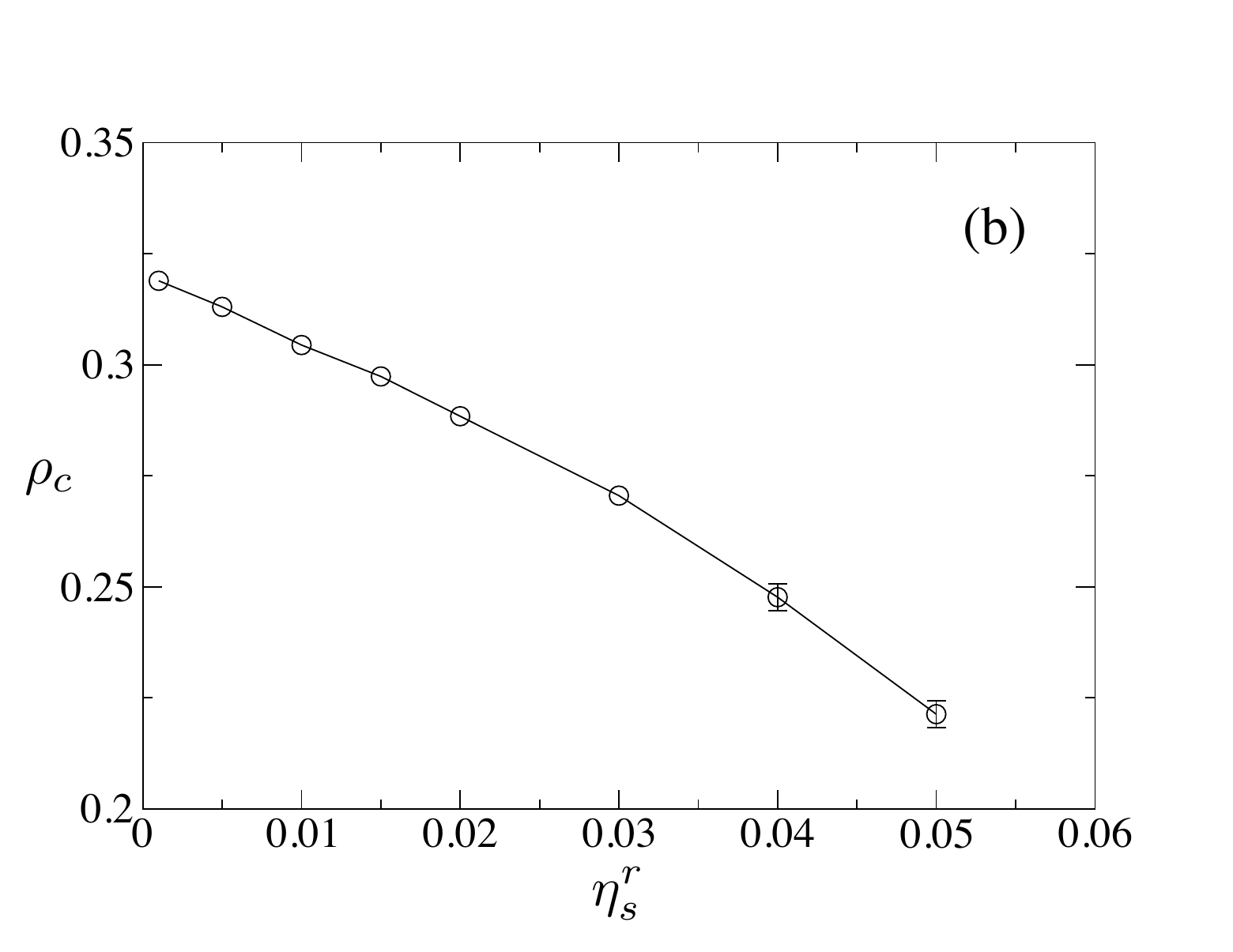} 
  \caption{(a)~Critical temperature and (b)~critical density versus $\eta_s^r$
    as determined from the iso-$Q^\star$ curves.  Error bars derive from a
    bootstrap analysis with 100 resamples.}
  \label{fig:critparams}
\end{figure}

\begin{figure}
  \includegraphics[width=\figurewidth]{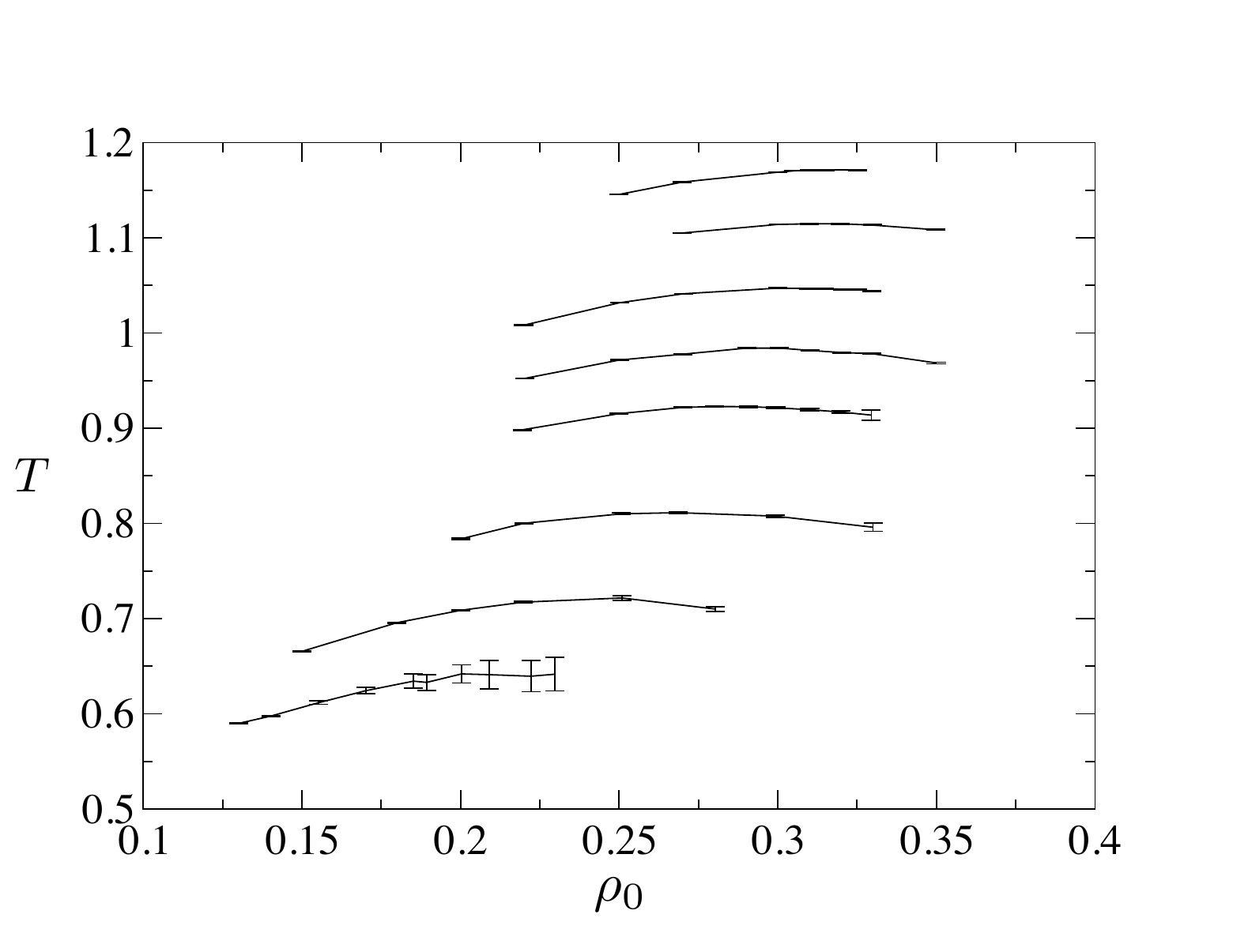} 
  \caption{The measured iso-$Q^\star$ curves for (top to bottom)
    $\eta^r_s=0$, $0.005$, $0.01$, $0.015$, $0.02$, $0.03$, $0.04$,
    and~$0.05$. Also shown are the estimated error bars from which we
    calculated the overall uncertainty in critical parameters via a bootstrap
    analysis.}
  \label{fig:allisoqstar}
\end{figure}

The critical-point parameters are determined, for each $\eta_s^r$ studied,
from measurements of the iso-$Q^\star$ curve.  For $\eta_s^r \le 3\%$, the
intersection method described in Sec.~\ref{sec:methods} is deployed to
determine the large-particle coexistence densities in the subcritical regime.
Figure~\ref{fig:phasediagram} presents our results for the $\rho$--$T$
binodal. The principal feature is -- as previously mentioned -- a strong depression of the binodal to lower
temperatures and lower densities as $\eta_s^r$ is increased. The scale of the
associated shifts in the critical parameters is made apparent in
Fig.~\ref{fig:critparams}, which plots our estimates of the critical
temperature and density as a function of $\eta_s^r$. One sees that as the
reservoir volume fraction of small particles is increased in the range
$0$ -- $0.05$, the critical temperature decreases by approximately $50\%$, while
the critical density drops by some~$30\%$. The error bars shown on the
estimates of the critical parameters derive from a bootstrap analysis of the
various quadratic fits that are consistent with the uncertainties in the locus
of each iso-$Q^\star$ curve (Fig.~\ref{fig:allisoqstar}).

\begin{figure} 
  \includegraphics[width=\figurewidth]{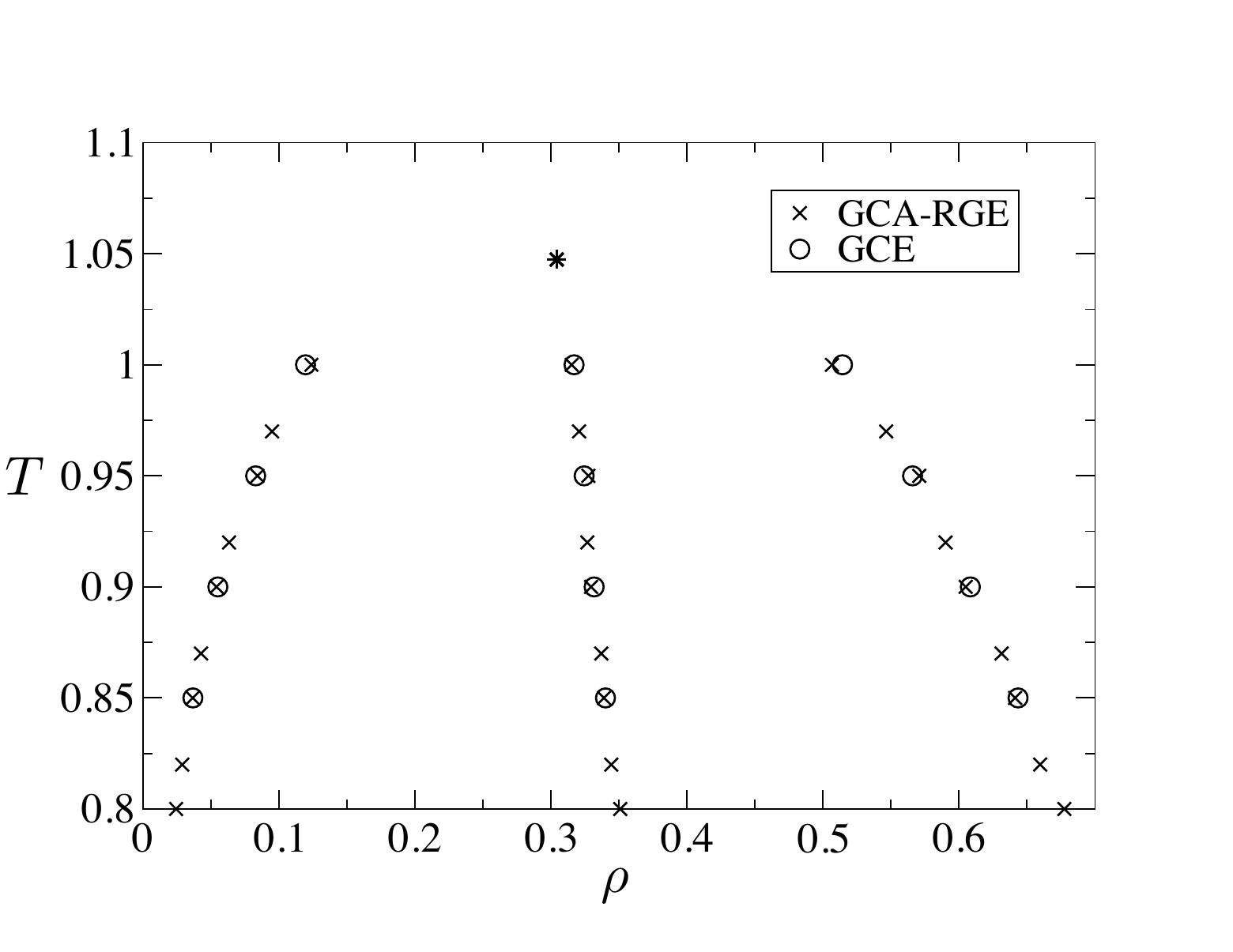} 
  \caption{Comparison for $\eta_r^s=0.01$ of the binodal obtained using the
    GCA--RGE technique (crosses) and the grand-canonical approach (circles)
    described in Ref.~\onlinecite{Ashton2010_A}. Statistical uncertainties are
comparable to the symbol sizes. Differences in the results
    near criticality arise from the different system sizes used in the two
    cases.}
  \label{fig:compare}   
\end{figure}

To demonstrate the correctness of our method, we compare the binodal for
$\eta_s^r=0.01$ with that obtained using a quite different approach, recently
proposed by two of us\cite{Ashton2010_A}.  This is a fully grand-canonical MC
scheme in which large particles are \emph{gradually} transferred to and from
the system by means of staged insertions and deletions. To negate ensemble
differences that occur when comparing results in finite-size systems, we
transform the grand-canonical distribution of the large-particle density,
$P(\rho)$, to the RGE using the exact transformation
$\hat{P}(\rho)=P(\rho)P(2\rho_0-\rho)$\cite{LIU2006,Ashton2010}.  We then
proceed to locate coexistence as if the data had been generated in the RGE by
treating $\rho_0$ as a parameter of the transformation.  The resulting
coexistence densities are compared with those obtained via the GCA--RGE
simulations in Fig.~\ref{fig:compare}. The agreement is good, particularly at
low temperature. The deviations near criticality arise from the difference in
the system size used in each case ($L=7.5\sigma_{ll}$ for the grand-canonical
system and $L=10\sigma_{ll}$ for the RGE system), and thus reflect that the
correlation length exceeds the system size in the grand-canonical simulation.

\begin{figure}
  \includegraphics[width=\figurewidth]{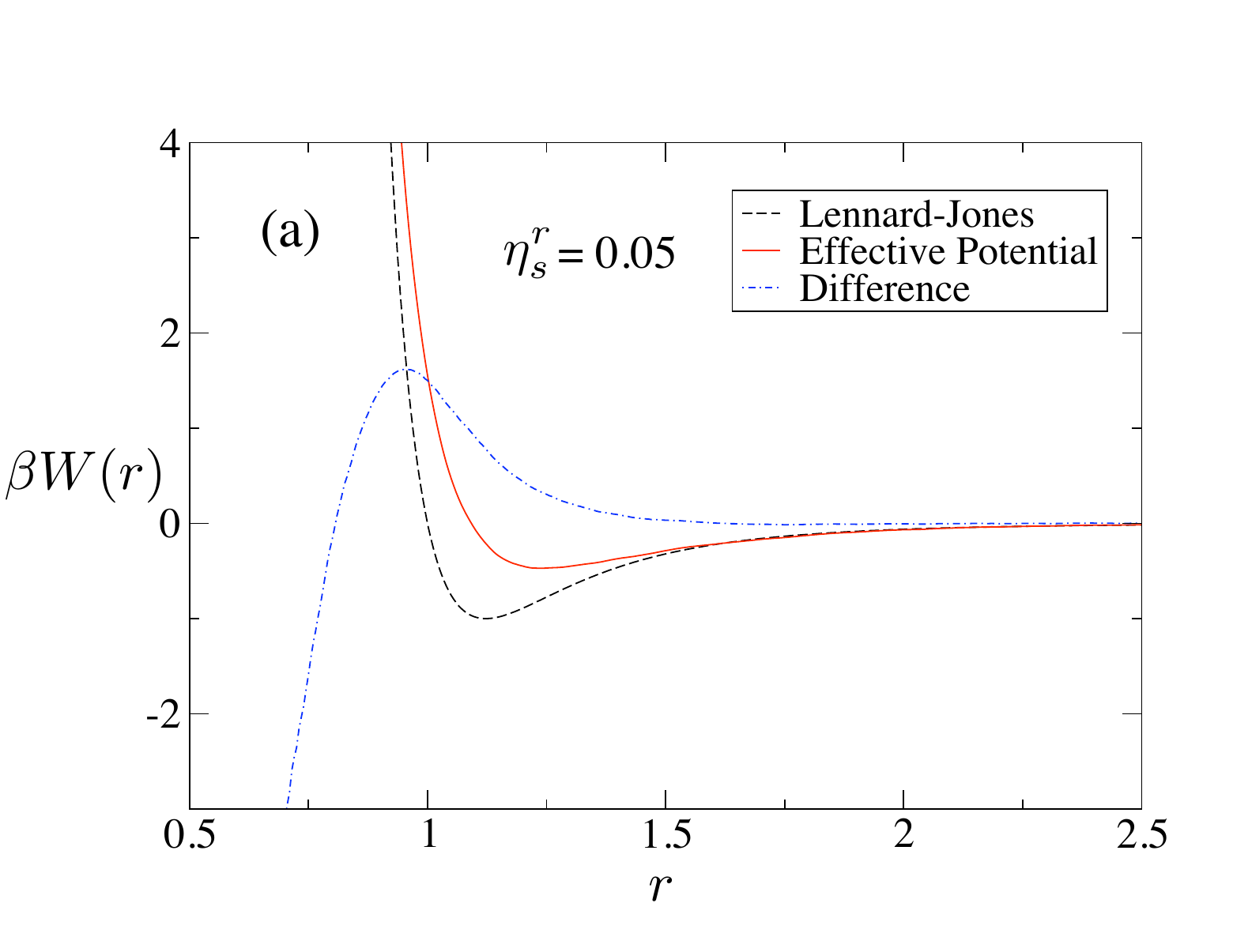} 
  \includegraphics[width=\figurewidth]{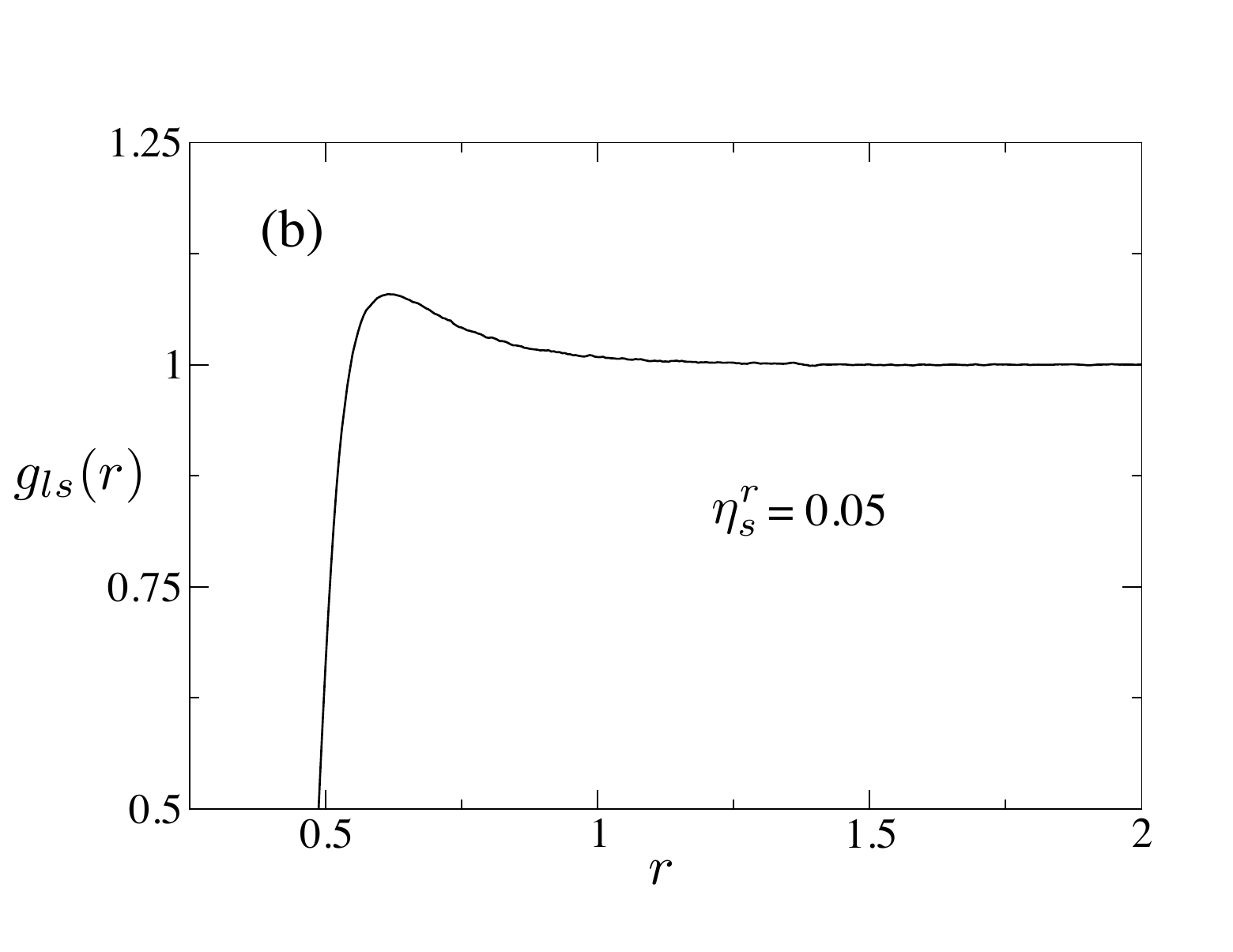} 
  \caption{(a)~The measured form of the effective potential $\beta W(r)$
defined in the text, at temperature $T=1.3$. Data are shown for the bare
LJ potential ($\eta_s^r=0$, dashed line) and $\eta_s^r=0.05$ (solid line) and their difference (dashed-dotted line).
(b) Form of $g_{ls}(r)$ for $\rho_{ll}\to 0$ at $\eta_s^r=0.05, T=1.3$.}
\label{fig:grs}  
\end{figure}

It is instructive to attempt to relate the shift in the binodal
occurring with increasing small-particle density to alterations in the
underlying local fluid structure.  An indication as to the factors at
work here follows from a study of the effect of small particles on the
effective potential between a pair of large particles,  \begin{equation}
\beta W(r)\equiv\lim_{\rho\to 0}-\ln[g_{ll}(r)]\;. \end{equation} An
example is shown in Fig.~\ref{fig:grs}(a) which compares $\beta W(r)$
for the cases $\eta_s^r=0$ (which simply corresponds to the bare
Lennard-Jones potential) with that for $\eta_s^r=0.05$, at $T=1.3$. One
sees  that for typical separations of large particles, the effective
potential is less attractive than the bare interaction. Thus the net
effect of the small particles is {\em repulsive} as shown by the
difference plot in fig.~\ref{fig:grs}(a), a feature that accords with
the reduction in the critical temperature. A likely reason for this is
to be found in the associated form of $g_{ls}(r)$ describing the
correlations between a large particle and a small particle, as shown in
Fig.~\ref{fig:grs}(b) at $\eta_s^r=0.05$. This shows that small
particles form a diffuse, non absorbing cloud around each large particle
because of their weak mutual attraction. Presumably, however, the
free-energy cost arising when the clouds associated with two or more
large particles overlap acts to reduce the intrinsic attractions between
large particles. Interestingly, the difference plot of
Fig.~\ref{fig:grs}(a) shows that at very small separations of large
particles (corresponding to high overlap energy) the effect of the small
particles changes from being repulsive to being attractive.

\begin{figure}
  \includegraphics[width=\figurewidth]{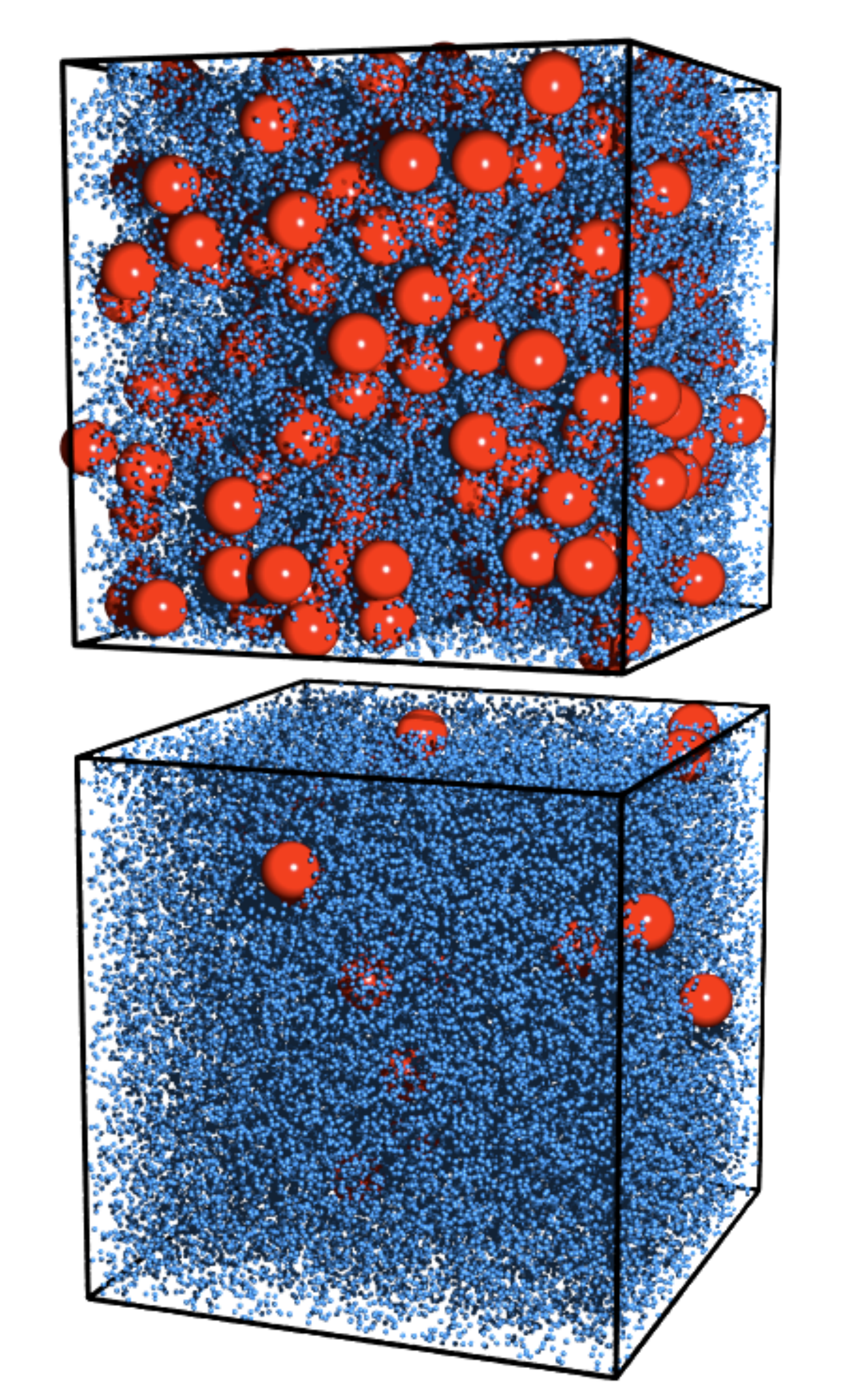} 
  \caption{Configurational snapshot of the two boxes in the restricted Gibbs
    ensemble at coexistence for $\eta^r_s=3\%$, $T=0.88T_c$.}
  \label{fig:snapshots}
\end{figure}

Finally, we show in Fig.~\ref{fig:snapshots} a configurational snapshot of our
simulation boxes at coexistence (i.e., $\rho_0=\rho_d$) for the case
$\eta^r_s=3\%$, $T=0.88T_c$. This provides a visual impression of the
character of the coexisting phases and the extent to which the large particles
are severely ``jammed'' by the small ones.

\section{Discussion and conclusions}
\label{sec:discuss}

In summary, we have described a variant of the Geometric Cluster
Algorithm\cite{LIU2004_0} for the accurate determination of phase behavior in
highly size-asymmetrical fluid mixtures. The method (an early version of which
was previously described in Ref.~\onlinecite{LIU2006}) operates by swapping
clusters containing large and small particles between two boxes of equal
volume, the global density of large particles being fixed. The resulting
spectrum of single-box fluctuations of the large particles can be analyzed
with respect to changes in their global density using the intersection method
of Ashton \emph{et al.}\cite{Ashton2010} to yield accurate estimates of
coexistence densities. Critical points can similarly be located to high
precision by using an appropriate finite-size estimator for criticality,
namely the maximum of the iso-$Q^\star$ curve.

We have applied the method to a LJ mixture with size ratio $10$:$1$ to
determine the coexistence properties of large particles for small-particle
reservoir volume fractions in the range $0 \le \eta_s^r \leq 3\%$.
Additionally, critical-point parameters were determined for $\eta_s^r=0.04$
and~$0.05$.  Our results show that when the small particles are weakly
attracted to the large ones, their net effect is to lower the degree of
attraction between large particles. As a consequence, the coexistence binodal
shifts to lower temperatures, confirming the preliminary findings of
Ref.~\onlinecite{LIU2006}.  Such a situation contrasts markedly with the
depletion effect applicable to small particles that interact with the large
ones like hard spheres\cite{Asakura1954}, for which there is a net increase in
the degree of attraction between large particles. Our measurements of local
structure suggest that in the case we have considered, the small particles
form a diffuse (nonadsorbing) cloud surrounding each large particle. The
overlap of clouds necessary for two large particles to approach one another
appears unfavorable in free-energy terms, leading to a net decrease in the
degree of attraction between large particles.  This is reminiscent of the
``nanoparticle haloing'' effect\cite{Tohver01,LIU2004_1}.

It is gratifying to note that the GCA--RGE method permits the study of
phase behavior in regimes that are inaccessible to traditional
simulation approaches. Specifically, the phase diagrams we have
presented could not have been obtained using even the most efficient
traditional approach to fluid phase equilibria, namely standard
grand-canonical simulation\cite{wilding1995}. For instance, for
$\eta_s^r>0.005$ our tests show that the grand-canonical relaxation time
is too large to be reliably estimated. Nevertheless, a lower bound on
the grand-canonical relaxation time, relative to that of the pure LJ
fluid, can be estimated via a comparison of the large-particle transfer
(insertion/deletion) acceptance probability~$p_{\rm acc}$. For
liquid-like densities of the large particles ($\rho\approx
0.6$), $p_{\rm acc}$ is of the order of $10^{-4}$ at
$\eta_s^r=0.005$\cite{LIU2006}.  Upon increasing $\eta_s^r$ to a volume
fraction of merely 1\%, this probability falls to $p_{\rm acc}\sim
10^{-6}$. These values are to be compared with $p_{\rm
  acc}\sim 10^{-1}$ for the pure LJ fluid. One can therefore expect the
grand-canonical relaxation time of the mixtures studied here to be several
orders of magnitude greater than for the pure LJ fluid. 

Notwithstanding the efficiency gains provided by the GCA-RGE approach,
it should be stressed that the results we have reported nevertheless
entailed a significant computational outlay. Specifically, runs to
determine each coexistence point typically varied in length between 100
and 3,000 hours of CPU time on a 3 GHz processor. The upper value in this
range was that required at the highest volume fraction of small
particles studied (for which there are very many small particles) and
the lowest temperature (where most of the large particles are involved
in each cluster update). Thus whilst studies of phase behaviour in
highly asymmetrical mixtures cannot yet be regarded as routine, they are
now at least feasible.

With regard to future studies of highly asymmetrical mixtures, one
barrier to attaining higher values of $\eta_s^r$ and smaller values of
$q$ is simply the computational overhead associated with large numbers
of small particles, although we note that the GCA has been successfully
applied to systems with millions of nanoparticles\cite{LIU2004_1}.
Additionally, in the present model, the suppression of the critical
temperature with increasing $\eta_s^r$ leads to a rapid growth in the
cluster size which renders the GCA--RGE approach increasingly less
efficient.  More generally, however, in situations where the small
particles induce an effective (depletion) attraction between the large
ones, we expect that the cluster size will remain manageable to rather
larger $\eta_s^r$ than studied here.

Finally, we mention an alternative approach, proposed by two of us, for
determining coexistence properties in highly size-asymmetrical
mixtures\cite{Ashton2010_A}.  This method utilizes an expanded
grand-canonical ensemble in which the insertion and deletion of large
particles is accomplished \emph{gradually} by traversing a series of
states in which a large particle interacts only partially with the
environment of small particles.  Implementing this approach requires
prior determination of a multicanonical weight function to bias
insertions of the particles, and thus renders it less straightforward to
use than the GCA--RGE. However, being fully grand canonical does have
the advantage of providing information on the chemical potentials of
large particles, thereby permitting histogram reweighting in terms of
density as well as temperature. In future work we hope to provide a
systematic comparison of the relative computational cost of both
approaches in various parameter regimes.

\acknowledgments 

This work was supported by EPSRC grant EP/F047800~(NBW) and NSF grant
DMR-0346914~(EL).  Computational results were partly produced on a machine
funded by HEFCE's Strategic Research Infrastructure fund.

\end{document}